\documentclass[reprint,amsmath,amssymb,aps,pra,longbibliography]{revtex4-1}
\usepackage{ulem}
\usepackage{graphicx}
\usepackage{dcolumn}
\usepackage{bm}
\usepackage{braket}
\usepackage[usenames, dvipsnames]{color}
\usepackage{lineno}

\begin{document}

\title{Experimental evidence of robust acoustic valley Hall edge states \\ 
	in a non-resonant topological elastic waveguide}

\author{Ting-Wei Liu}
\author{Fabio Semperlotti}
 \email{fsemperl@purdue.edu}
\affiliation{Ray W. Herrick Laboratories, School of Mechanical Engineering, Purdue University, West Lafayette, Indiana 47907, USA.}

\begin{abstract}
This paper presents experimental evidence of the existence of acoustic valley Hall (AVHE) edge states in topological elastic waveguides. The fundamental lattice is assembled based on a non-resonant unit where space inversion symmetry (SIS) is broken by simply perturbing the underlying lattice geometry. This aspect is in net contrast with existing elastic AVHE designs that exploit locally-resonant units and require the addition of masses in order to break SIS. The experimental results presented in this study validate findings so far presented only at theoretical and numerical level. In particular, it is found that edge modes can effectively propagate along domain walls between topologically dissimilar domains and that disorder-induced backscattering is substantially suppressed due to the weak coupling between oppositely valley-polarized modes. The coupling between valley modes is also further investigated and linked to an evident chiral flux of the mechanical energy. Finally, we show that the weak coupling between the valleys can be exploited to achieve selective mode injection at the domain wall, hence realizing a very effective excitation strategy of the chiral edge states.
\end{abstract}

\maketitle

\section{introduction}
Controlling the flux of acoustic energy in elastic structures has been a long-standing challenge and an active area of research in engineering. Almost endless are the examples or real-world applications that could greatly benefit from design methodologies capable of synthesizing materials and structures having a high level of control on the propagation of acoustic and elastic waves. Examples range from efficient on-chip devices (such as microfluidic manipulation \cite{droplet,mix}, cell sorting \cite{tumor} on biochips, and signal processing in mobile communication devices \cite{YY,AT,ST}), to vibration and noise control in mechanical systems
\cite{SEMP1, SEMP2, SEMP5, SEMP6, SEMP7, SEMP8, sun,lin,TT}, to vibration-based energy harvesting \cite{SEMP3}. Scientists have long been exploring approaches to design waveguides operating in the scattering regime and capable of guiding acoustic (or elastic) waves around sharp impedance discontinuities (e.g. corners or structural defects), without giving rise to appreciable backscattering.

Some novel directions came in recent years from the field of topological acoustics. Owing to the many similarities (both at mathematical and physical level) between acoustics and the wave nature of electrons, researchers have recently leveraged the groundbreaking discoveries of topological phases of matter \cite{ReviewKaneTI, ReviewNiu} and created their acoustic analogues \cite{TopoPhon-Gyro, TopoGyroExp, TopologicalAcoustics-Flow, TopoSound-Flow, TunableTopoPnc-Flow, MechanicalTI,  AcousticTIPlate, AcousticTIAir, pspin, ValleySonicBulk,  ValleySonicEdge, ValleySonicEdge, Ruzzene1, Ruzzene2, myqvhe, Bistable, diatom, valleyvein}. Among the most significant properties of topological materials, there is certainly the ability to induce chiral and backscattering-immune edge states supported at the boundaries of topologically different materials.

The first implementation of a topological acoustic material was designed after a mechanism analogue to the quantum Hall effect (QHE). In this material, spinning rotors \cite{ TopoPhon-Gyro, TopoGyroExp} or circulating fluids \cite{TopologicalAcoustics-Flow, TopoSound-Flow, TunableTopoPnc-Flow} were embedded within the supporting acoustic or elastic waveguide in order to break time-reversal symmetry (TRS). Although these designs were able to achieve nonreciprocal edge states propagating unidirectionally, their fabrication complexity made them less viable for practical applications. Later studies concentrated on exploiting the acoustic analogue of the quantum spin Hall effect (QSHE) \cite{MechanicalTI, AcousticTIPlate, AcousticTIAir, pspin} which, contrarily to the QHE, does not require TRS breaking. These initial studies were able to show that, by creating acoustic pseudo-spins and pseudo-spin-dependent effective fields, edge states topologically protected from back-scattering could be successfully achieved. 

Only in very recent years, researchers have explored the possibility to design both acoustic and elastic topological waveguides by exploiting the acoustic analogue of the quantum valley Hall effect (QVHE; aoustic valley Hall effect, AVHE) \cite{ValleySonicBulk, ValleySonicEdge, Ruzzene1, Ruzzene2, myqvhe, Bistable, diatom, valleyvein,stonely,snowflake,2018avhe}.
The underlying mechanism of the QVHE/AVHE requires only space-inversion symmetry (SIS) breaking in a lattice that possesses Dirac dispersion at the high symmetry points. This condition can be realized fairly easily in an artificial metamaterial. In electronic materials, SIS was broken using different methodologies including graphene-like lattices with staggered sublattice potentials \cite{ValleyContrasting, DomainWall, ControlEdgeStates, SiliceneBrokenSIS,SiliceneDW}, strained lattices \cite{StrainedGraphene}, and multi-layered graphene under electric fields \cite{EdgeStatesMultilayerGraphene,MultilayerGraphene,BilayerGrapheneDW, ValleyChernBilayerGraphene,BilayerGraphene,Highway,GateControl}. The effect of the SIS breaking was to open a topological bandgap between the Dirac cones associated with the \textbf{K} and $\mathbf{K'}$ symmetry points, hence creating the conditions for the existence of edge states. These edge states could not be explained in the context of either QHE or QSHE. 
In fact, the lattice still possesses a trivial topology within the context of QHE \cite{ReviewKaneTI, Raghu, PhotonicGraphene} because TRS is intact, and it also possesses a trivial topology from a QSHE perspective because it lacks the spin degree of freedom. However, due to the large separation in $\mathbf{k}$-space of the two valleys \cite{ando1,ando2} and to the localized distribution of non-zero Berry curvature, a valley-dependent topological invariant can be defined and used to classify the topological states of the different lattices. 

It is well-known that in a 2D hexagonal lattice structure, having both TRS and SIS intact, the band structure has deterministic degeneracies occurring at the Brillouin zone (BZ) corners (also called valleys), namely the $\mathbf{K}$ and $\mathbf{K}'$ points. In the neighborhood of these points, the band structure exhibits linear and isotropic dispersion with a typical conical profile. Owing to their characteristic geometric shape and to the fact that around these degeneracies the system dynamics maps to the massless Dirac equations \cite{myqvhe,diatom}, these dispersion structures are often referred to as Dirac cones (DC). The Berry curvature associated to the upper and lower branches forming the DC is null everywhere in \textbf{k}-space except in the neighborhood of the degenerate $\mathbf{K}$ and $\mathbf{K}'$ points where it is undetermined. Under these conditions, the Chern number (a topological invariant obtained integrating the Berry curvature around the complete Brillouin zone) is also identically zero, hence confirming the topologically trivial nature of the original lattice. 

When SIS is broken due to a perturbation of the original lattice, the degeneracies are lifted and a band gap opens up at the original Dirac points. This condition is associated with two major changes. First, the Berry curvature becomes a smooth function that peaks around each valley. Second, the valley Chern number (that is the Chern number calculated in a finite area around the valley, not over the entire BZ) returns non-zero values, hence suggesting a (local) topological significance of the band structure \cite{ValleyContrasting, myqvhe}. 
When lattices having different valley Chern numbers are assembled together, topological edge states can exist at the domain wall (DW), that is the physical interface between topologically distinct phases. These edge states are characterized by a strong (although not complete) chirality which results in partial suppression of defect-induced backscattering. This connection between the topological nature of the bulk and the occurrence of edge modes at the boundaries where the topological transition occurs, is often referred to as the \textit{bulk-edge correspondence}.

The acoustic version of the valley Hall effect was recently investigated for application to mechanical systems. Lu \textit{et al.} \cite{ValleySonicBulk, ValleySonicEdge} proposed a fluidic acoustic design. Later, Pal \cite{Ruzzene1} and Vila \cite{Ruzzene2} theoretically and experimentally showed the AVHE edge state in a tight-binding like (locally-resonant) elastic waveguide. Liu \textit{et al.} \cite{myqvhe} used a fully continuum approach to study AVHE of non-resonant elastic waveguide with different symmetry, and also proposed a method to excite specific unidirectional edge states on the domain wall. Following the approach of this latter study, this paper presents experimental evidence of the existence of AVHE edge states in a non-resonant phononic thin lattice, as well as it validates the corresponding unidirectional edge state excitation. In addition, this paper also presents an interpretation of the AVHE in terms of energy vortices that bears some interesting, although purely qualitative, similarity with the role of spin in QSHE.

More specifically, this work provides several new contributions with respect to the very limited experimental literature available on elastic AVHE waveguides \cite{Ruzzene2,diatom}.
First of all, it should be highlighted that the only experimental evidence of elastic AVHE deals with locally-resonant material systems. This result is a natural consequence of the fact that the theoretical background of AVHE was originally developed in electronic systems based on the tight-binding approximation. Such approximation assumes that electrons localize around the atom sites, which is a good assumption for material systems like graphene and even for locally-resonant phononic structures, given that the eigenstates tend to localize near each resonator. However, the fundamental tight-binding assumption breaks down for non-resonant phononic structures, therefore making the extension of AVHE to non-resonant materials certainly not a foregone conclusion. By using theoretical arguments as well as semi-analytical and numerical methods, Liu et al. \cite{myqvhe} showed evidence that non-resonant phononic crystals could indeed exhibit AVHE topological behavior. However, an experimental validation of these findings was still lacking in the literature.
Although previous experimental works from Vila et al. \cite{Ruzzene2} and Zhu et al. \cite{diatom} performed analysis of elastic AVHE without employing a tight-binding approximation, the phononic lattices in those studies were characterized by strong spatial modulations in mechanical impedance (either via added masses or carved notches, or both). Such strong changes in local impedance makes the acoustic wavefunction strongly localized around the ``atom'' site. This is a feature that is reminiscent of resonant characteristics. In fact, such localization is analogous to what happens to electronic wavefunctions near the nuclear potential and it is a key factor for the application of a tight-binding approach. On the contrary,
the present study provides definitive experimental evidence that elastic valley Hall edge states can also be realized in a non-resonant type phononic crystal,
generalizing the application to a wider range of phononic lattice designs beyond the resonant category.

The use of a non-resonant design has important implications for the development of tunable and reconfigurable materials. In their previous theoretical work \cite{myqvhe}, the authors discussed the possibility to realize real-time tuning by applying internal pressure loading. The numerical predictions suggested that the nonlinear geometric effects associated with the load-induced prestress were negligible and did not affect, modify, or hinder the topological behavior in a significant way. In the present work, the experimental setup was built around these theoretical observations therefore producing SIS breaking not via the application of an external action but by fabricating the lattice already in deformed configuration. The experimental results confirm that the effect of the prestress and of the geometric nonlinearities (stiffening) due to the elastic deformation of the unit cell are indeed negligible and do not alter the performance of the topological material. This result not only confirms previous theoretical predictions but, more importantly, suggests that actively-controlled elastic deformations are a very suitable option to reconfigure and tune topological materials.

Another relevant contribution and notable departure compared to previous experimental works \cite{Ruzzene2,diatom} consists in the underlying mechanism used to break SIS. Previous studies relied on adding or removing localized masses at specific lattice sites in order to break the space inversion symmetry. This was a direct transposition to mechanical systems of the concept of staggered sublattice potentials used in graphene. In mechanical systems, this aspect has two main implications: 1)	the locally-resonant elements have an inertia-driven response which is in net contrast with our design which is instead characterized by a controlled-distribution (via unit cell geometry) of the local stiffness, and 2) the added masses permanently break SIS hence providing a design not suitable for real-time control and reconfiguration of the material. In our proposed design, the lattice in its undeformed state loses any topological property, hence creating a more versatile and reconfigurable system that could potentially switch on and off between different topological states.
This paper also presents experimental evidence of a simple yet effective procedure capable of achieving selective excitation of the unidirectional edge states. The approach is based on a two-transducer setup that was presented and numerically evaluated in our previous study \cite{myqvhe}.
Finally, we report for the first time the numerical and experimental observation of vortices of mechanical energy flux/phase distribution in valley-polarized modes. We use this physical observation to provide an intuitive, although more qualitative, physical insight in the nature of chiral edge modes and of their back-scattering immunity in AVHE.

\section{Topological elastic waveguide}

\begin{figure}[h]
	\includegraphics[scale=.95]{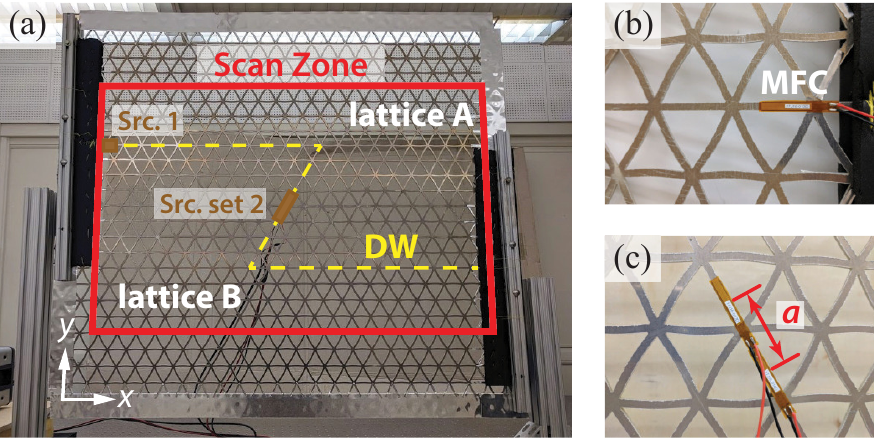}
	\caption{\label{fig:setup} Experimental setup. (a) Front view of the mesh-like aluminum thin plate with hexagonal lattice structure and $D_{3h}$ symmetry. The plate is partitioned in two subdomains characterized by broken SIS. The two domains are an inverted version of each other and are labeled lattices A and B. The interface between the two lattices is the domain wall (DW), indicated by the yellow dashed line. The two brown rectangles labeled Src.~1 and Src.~set 2 show the location where MFC actuators are bonded on the plate (back side). (b) and (c) show a back side view of the plate where Src.~1 and Src.~set 2 are located.}
\end{figure}

The test structure considered in this study consists in a reticular aluminum thin plate having hexagonal lattice structure, as shown in Fig.~\ref{fig:setup} (a). The reticular plate was obtained by cutting the lattice from a flat aluminum sheet of thickness 0.08 inches. Note that, as explained above, the fabricated plate already represents the SIS-broken lattice (the individual elements already possess a slight curvature). In this configuration, the lattice has $D_{3h}$ symmetry as opposed to the undeformed lattice (i.e. the one having straight trusses) which has $D_{6h}$ symmetry and SIS fully preserved. 

The panel is divided into two different subdomains characterized by inverted geometric patterns and denominated lattice A and B (Fig.~\ref{fig:setup} (a)). The interface between the lattices forms a DW consisting of flat trusses (indicated by the yellow dashed line in Fig.~\ref{fig:setup} (a)). The geometric patterns of the unit cells associated with lattices A and B are shown in Fig.~\ref{fig:bs} (a) and (b), respectively. The individual truss elements has width $w=0.1a$ and thickness $h=0.05a$, and the lattice constant is $a=40.64\mbox{ mm}$.

\subsection{Topological band structure analysis}

\begin{figure}[h]
	\includegraphics[scale=1.5]{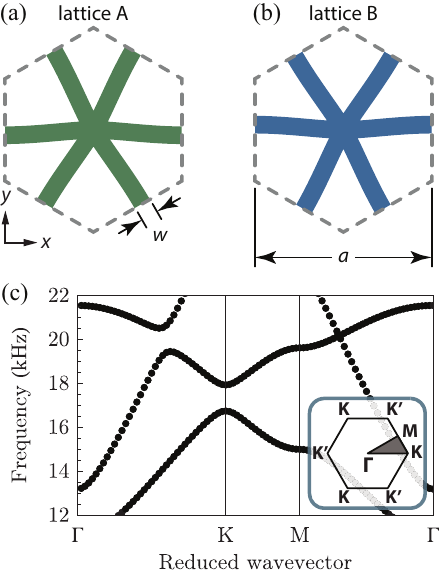}
	\caption{\label{fig:bs} (a, b) The geometric patterns of the fundamental unit cells of the lattice A and B. The images shows the deformed lattice structure used to induce SIS breaking. The two lattices are inverted images of each other hence they share the same (c) dispersion band structure. A partial band gap opens at the \textbf{K} point as a result of SIS breaking. The inset shows the symmetry points and the irreducible Brillouin zone (grey triangle).}
\end{figure}

Given that TRS is intact, the two mirror lattices share identical band structures. Fig.~\ref{fig:bs} (c) shows the band structure limited to the flexural antisymmetric Lamb modes. The SIS breaking induced by the perturbation of the initial hexagonal geometry lifts the degeneracy at the \textbf{K} point and leaves behind a topological bandgap. To further clarify this point, consider a continuous and smooth deformation, for example, of lattice A that gradually morphs it into lattice B. During this continuous transition the bandgap closes (when the lattice matches the undeformed reference pattern having intact SIS) and then it reopens again when the deformation towards lattice B begins. This evolution suggests that a topological phase transition could be taking place when the bandgap closes and reopens. In fact, previous studies showed that by using a semi-analytical $\mathbf{k}\cdot\mathbf{p}$ approach \cite{myqvhe}, the phononic dispersion around each valley of such lattice can be described by the massive Dirac equation and the associated valley Chern numbers switch from $C_{\nu,{\mathbf K}}^\mathrm{A,up}=-1/2$ (the valley Chern numbers of the upper mode of lattice A at the valley \textbf{K}) to $C_{\nu,{\mathbf K}}^\mathrm{B,up}=+1/2$.

Equivalently, switching to the other valley also switches the sign of the valley Chern number; a direct consequence of time reversal symmetry (TRS).
In addition, the Berry curvature of the lower mode carries an opposite sign compared to its upper companion. This observation is consistent with the fact that the sum of the Berry curvatures of all modes must vanish at a given wavevector.

\subsection{Topological edge states at the domain wall}

\begin{figure*}[ht]
	\includegraphics[scale=0.85]{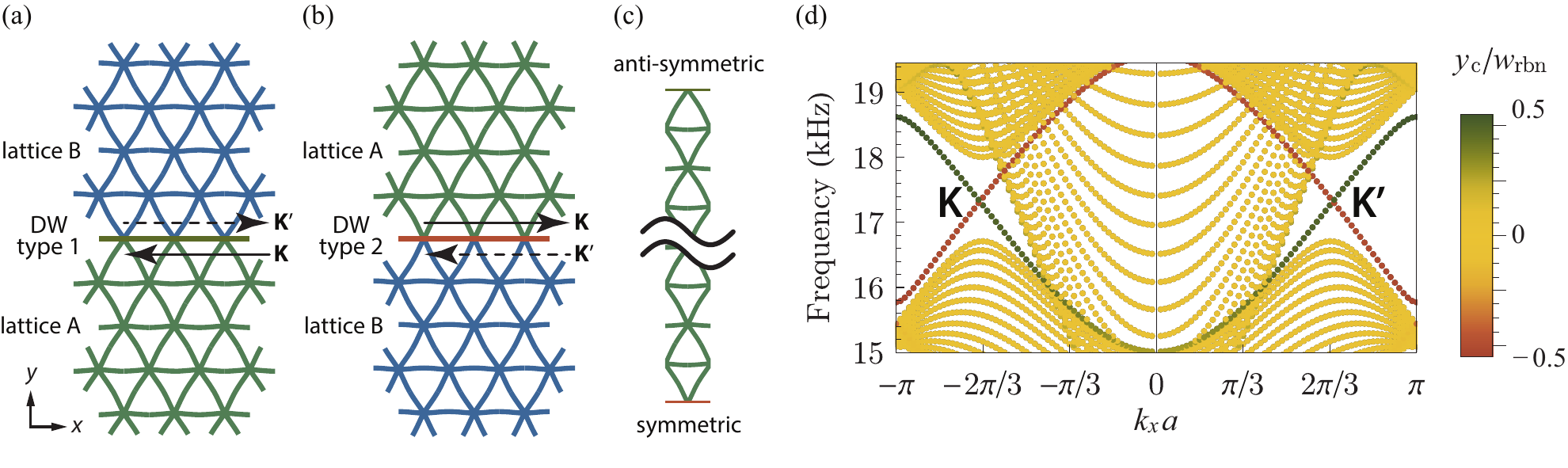}
	\caption{\label{fig:DW} (a) Illustrations of type-1 DW assembled by connecting lattice B above A.  (b) Type-2 DW assembled by connecting lattice A above B. (c) The superlattice used in calculating the edge state dispersion. It is composed of lattice B, bonded in positive and negative $y$-direction with type-1 and type-2 DW configurations accompanied by symmetric and antisymmetric boundary conditions, respectively. (d) The dispersion of the superlattice. The colorbar denotes the $y$-position of the center of strain energy distribution ($y_\mathrm c$) normalized by the ribbon width ($w_\mathrm{rbn}$) in order to emphasize the edge states. Modes in green indicate edge states at type-1 DW, while modes in red indicate edge states at type-2 DW. The valley index is also indicated.}
\end{figure*}

According to the concept of bulk-edge correspondence \citep{ValleyContrasting}, one gapless edge state should be expected at the DW between lattices A and B. In order to confirm this observation, we performed a numerical supercell analysis to determine the edge states dispersion and the corresponding eigenmodes. Depending on the nature of the interface between the two lattices, two types of DWs can be achieved by assembling either lattice B above A or viceversa. The resulting domain walls are labeled type 1 and type 2, respectively (see Fig.~\ref{fig:DW} (a), (b)).

Considering that the DWs are planes of mirror symmetry, the edge states can be either symmetric or antisymmetric with respect to the DW center line. In our calculations, we built a finite ribbon (Fig.~\ref{fig:DW} (c)) composed of lattice B, bonded in both the positive and negative $y$-directions with both type-1 and type-2 DW configurations (in addition to either symmetric or antisymmetric boundary conditions). This modeling choice was made in order to capture all the admissible edge states in a single model.
 
The edge states at type-1 and type-2 DWs were then calculated and included in the overall band structure (green and red dotted lines in Fig.~\ref{fig:DW}, respectively). The color scheme in Fig.~\ref{fig:DW}(b) represents the $y$-position of the centroid of the strain energy distribution ($y_\mathrm c$), normalized by the ribbon width $w_\mathrm{rbn}$, such that $y_\mathrm c / w_\mathrm{rbn} = \pm 0.5$ indicates the ribbon's edge (or equivalently the two DWs). By inspecting the dispersion curves, it is clear that the modes crossing the topological bandgap are confined at the edges of the ribbon. Note that although there is no complete band gap for the bulk modes, the edge states still possess a non-leaky nature as long as the bulk modes are faster than the edge states at the same frequency \cite{myqvhe}.

It is worth highlighting that, independently of the type of DW, every edge state has its own time-reversal (TR) counter part. 
This means that both right and left traveling edge states exist at the same DW, but have wave numbers corresponding to the two distinct valleys (see Fig.~\ref{fig:DW} (d) and arrows in (a) and (b)).
In principle, this would suggest that unidirectional propagation could be achieved on these DWs even when in presence of localized disorder, defects of the lattice, or discontinuities. In practice, as the discontinuity does not induce appreciable intervalley hopping (i.e. the process of steering the wavevector towards the opposite valley after scattering) which would require a large crystal momentum change, reflections are minimized and quasi-uni-directional propagation is effectively achieved.

\begin{figure}[h]
	\includegraphics[scale=1.4]{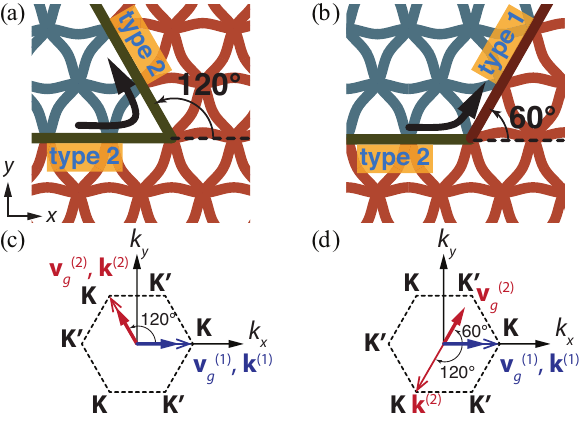}
	\caption{\label{fig:type} (a) A $60^\circ$ acute DW corner connects DWs of the same type. (b) A $120^\circ$ DW corner is always accompanied by a DW type switching.}
\end{figure}

To understand the behavior of a wave traveling through a discontinuity of the DW (e.g. a change in direction), we first investigate a $60^\circ$ acute corner made entirely of type-2 DWs (see Fig.~\ref{fig:type} (a)). Transmission through this corner requires a 120$^\circ$ deflection of the energy flow (or equivalently, of the group velocity vectors $\mathbf{v}_g^{(1)} \rightarrow \mathbf{v}_g^{(2)}$ as the thick solid arrows shown in Fig.~\ref{fig:type} (c)). We also note that on the first wall segment, the edge state with $+x$-aligned (i.e. right going) group velocity has a wavenumber $k_x=-2\pi/3a$ (see the red dots in the left half of Fig.~\ref{fig:DW} (d)). This wavevector is corresponding to the crystal momentum of the valley \textbf{K}, and it is schematically represented as the thin blue
 hollow arrow $\mathbf k^{(1)}$ in Fig.~\ref{fig:type} (c). Note that here the terminology \textit{crystal momentum} is used to refer to the periodic representation of the wavevector in the reciprocal space that is any two wavevectors differing by an integer number of reciprocal basis vectors are considered equivalent. In other terms, all the \textbf{K} points in the \textbf{k}-space indicate the same crystal momentum.
Once the wave propagates through the corner and it is transmitted to the second segment of the DW, the condition is described in an exactly equivalent way by simply rotating the momentum vectors of 120$^\circ$ (that is, $\mathbf k^{(2)}$ is the thin red hollow arrow in Fig.~\ref{fig:type} (c)). It is found that the crystal momenta of the edge states $\mathbf k^{(1)}$ and $\mathbf k^{(2)}$ are identical (to an added integer number of reciprocal lattice vectors) and they both point at the valley \textbf{K}. Such transmission involves no inter-valley hopping thus the edge state experiences no reflection at the sharp bent.

Consider now a $120^\circ$ corner on the DW ($60^\circ$ deflection of the energy flow as shown in Fig.~\ref{fig:type} (b)). In this case, we can show that intervalley hopping does not take place therefore strongly suppressing reflections from the corners. Due to lattice symmetry, a $120^\circ$ DW corner must be accompanied by a change in DW type, as shown in Fig.~\ref{fig:type} (b). The edge states associated with different types of DWs exhibit opposite group velocity, as already illustrated in Fig.~\ref{fig:DW} (d). Therefore, the group velocity vector is bent by 60$^\circ$ after passing through the corner, while the crystal momentum vector still points at the same valley. To clarify this condition, we consider a specific example in which the propagating edge state travels on a DW type-2 before the corner and on a DW type-1 after (see Fig.~\ref{fig:type} (b)). When the wave travels on the first wall segment (type-2), the situation is described in an analogous way to the previous case of a $60^\circ$ corner; $\mathbf{v}_g^{(1)}$ and $\mathbf{k}^{(1)}$ are indicated by the blue arrows in Fig.~\ref{fig:type} (d).
For the second segment (type-1), consider first a left-propagating edge state on DW type-1 parallel to the $x$-axis
(green dots in the left half of Fig.~\ref{fig:DW} (d)). We can draw the corresponding arrows $\mathbf{v}_g$ aligned with negative $x$-axis and the wavevector $\mathbf{k}$ pointing at the valley $\mathbf{K}$.
If we rotate clockwise by $120^\circ$ this figure, we obtain the DW of the second segment. The resulting lattice still has $C_3$ symmetry which means that the $120^\circ$ rotation does not change either the lattice or the DW type.
Now, draw $\mathbf{v}_g^{(2)}$ and $\mathbf{k}^{(2)}$ as the red arrows in Fig.~\ref{fig:type} (d). 
The initially negatively oriented $x$-axis group velocity after the $120^\circ$ clockwise rotation is shown by the red solid arrow $\mathbf{v}_g^{(2)}$. After the rotation, also the crystal momentum still points at \textbf{K}.
This observation further confirms that $\mathbf{k}^{(1)}$ and $\mathbf{k}^{(2)}$ point at the same valley and transmission through an 120$^\circ$-corner should be expected to be reflection-free.

\begin{figure}[h]
	\includegraphics[scale=0.85]{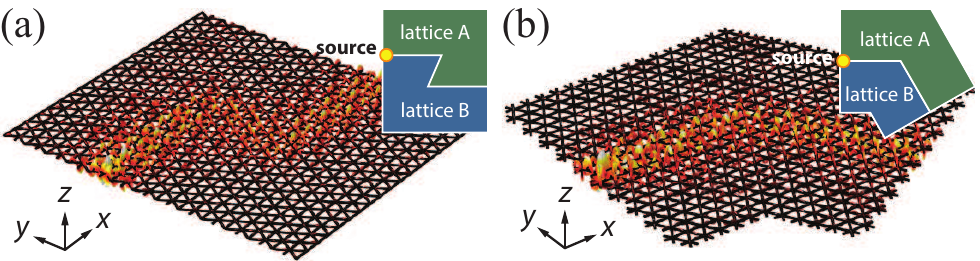}
	\caption{\label{fig:ZDW} Full field numerical simulations of edge state transmission through (a) a 60$^\circ$ and (b) a 120$^\circ$ corner. The insets show the lattice types and the shape of the DWs. The yellow dot indicates the position of the harmonic excitation source. Low-reflecting boundary conditions were applied all around to suppress unwanted reflections. Edge states concentrate at the DWs and have uniform amplitudes therefore suggesting no reflection at the corners.}
\end{figure}

Full field numerical simulations of edge state transmission through both 60$^\circ$ and 120$^\circ$ corners were carried out and shown in Fig.~\ref{fig:ZDW} (a) and (b), respectively. Edge states concentrate at the DWs and have uniform amplitudes throughout, therefore suggesting that no reflection occurs at the corners.
In these simulations, a point source was placed at one end of the domain wall, and low-reflecting boundary conditions were applied on the model boundaries to suppress unwanted reflections.

\begin{figure}[h]
	\includegraphics[scale=1]{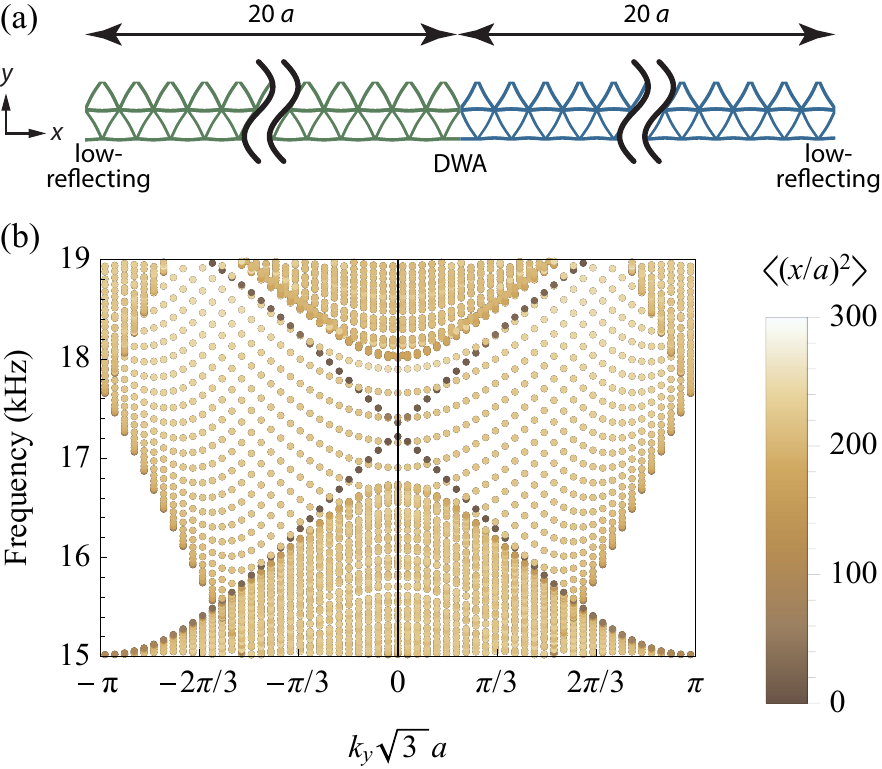}
	\caption{\label{fig:armchair} (a) The supercell with armchair type DW (denoted as DWA). Each lattice (A: green and B: blue) has width of $20a$. Low reflecting boundary conditions are applied on the left and right ends. The wave is propagating along $y$-direction. (b) The dispersion of the supercell. The color denotes the $x$ deviation weighted by $(\textrm{kinetic  energy}+\textrm{strain  energy})$. The edge state is topologically trivial and has a band gap.}
\end{figure}

In addition to zigzag types of DWs, the edge modes along armchair (parallel to $\Gamma$-M) type of DWs are also investigated. Fig.~\ref{fig:armchair} (b) shows the dispersions of the supercell with armchair type DW connection shown in Fig.~\ref{fig:armchair} (a). The supercell consists lattices A (green) and B (blue) connected by the armchair type interface denoted as DWA. And the elastic wave is assumed propagating in $y$-direction. Since DWA is parallel to $\Gamma$-M, the wavevector is never close to the valleys in the reciprocal space where the topological invariant is defined. In fact the wavevector is always equally distant from the two valleys $\mathbf{K}$ and $\mathbf{K}'$. Therefore considering wave propagating along $\Gamma$-M direction, there is no topological distinction between the two adjacent lattices A/B, and therefore no guaranteed protected edge states. On the other hand, the edge state and its TR counterpart meet near $k_y=0$. Contrasting to the valley edge states along zigzag DWs (see the green (or red) modes in Fig.~\ref{fig:DW} (d)) which they and their TR counterparts are far separated in the $k$-space, the closeness in wavevector and frequency plus no preferred valley polarization make back-scattering easily initiated by disorders. The small gap caused by repulsion between the two TR edge states indicates coupling between them. Similar observation is also reported by Yang \textit{et al.} \cite{2018avhe} in a fluidic acoustic system.

\section{experimental validation}
The experimental validation was divided in two different phases. In the first phase, we examined the existence and robustness of the edge states. In the second phase, we tested the selective valley injection procedure.

\subsection{Existence of the edge states}

In order to test the propagation conditions on the DWs, guided wave modes were generated using a single micro fiber composite (MFC) thin film transducer glued on the back of the panel (Fig.~\ref{fig:setup} (b)) at the location labeled Src.~1 (Fig.~\ref{fig:setup} (a)). The response of the plate was measured using a Polytec PSV-500 scanning laser Doppler vibrometer which provided the velocity field over the entire panel.
Viscoelastic damping tape (3M 2552) was also applied along the plate boundaries in order to reduce reflections.
Note that, without the tape, the domain wall terminal (i.e. the plate boundary) would give rise to strong reflection of the edge state. This observation is consistent with the fact that the QVHE is a weak topological material and enables back-scattering immune behavior only for so-called light disorder \citep{AllSi}.

\begin{figure}[h]
	\includegraphics[scale=0.85]{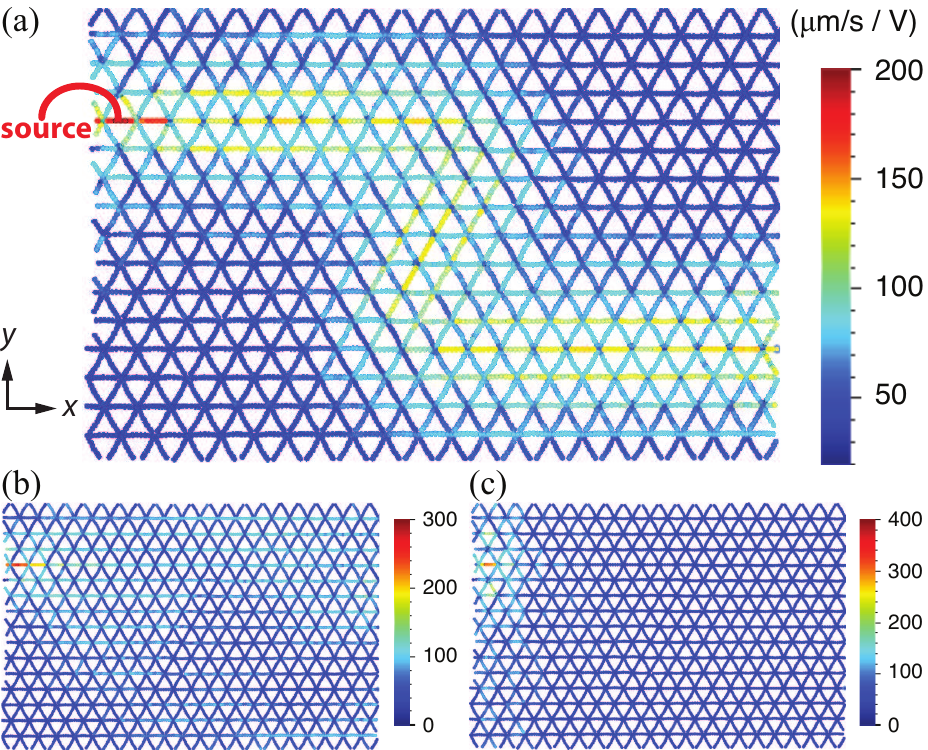}
	\caption{\label{fig:Z} (a) Averaged amplitude of the measured transfer function (velocity/input voltage) at steady state in the frequency band 16.9-18.3 kHz. The highly localized response clearly indicates the existence of the edge state at the Z-shaped DW. (b) The response at 16 kHz. (c) The response at 19 kHz.}
\end{figure}

Fig.~\ref{fig:Z}(a) shows the averaged amplitude of the measured transfer function (velocity/input voltage) at steady state in the frequency band 16.9-18.3 kHz, which is the range of existence of the edge states. The strongly localized response at the Z-shaped DW confirms the existence of the edge state. Although some bulk modes are supported in the same frequency range (Fig. \ref{fig:DW}(b)), their amplitude decays out quickly from the source ($\propto 1/r$) and becomes negligible compared to the edge states. On the other side, the edge state is non-leaky and maintains its amplitude as it propagates along the DW.
As a comparison, the responses around 16 kHz and 19 kHz are plotted in Fig.~\ref{fig:Z}(b) and (c) to further illustrate the frequency selectivity of the edge sate.

\begin{figure}[h]
	\includegraphics[scale=1]{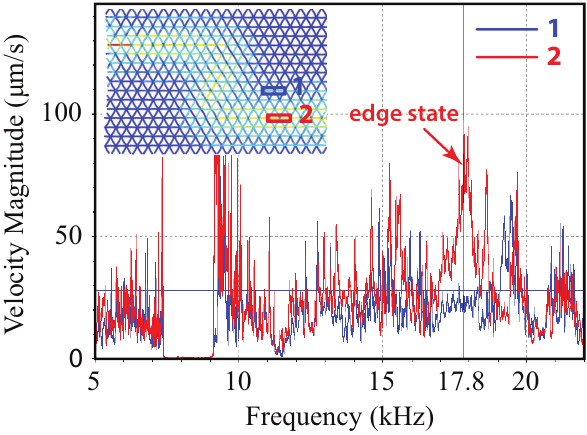}
	\caption{\label{fig:spectrum}The amplitude of the velocity spectra at two different locations on and off the edge state. The large difference between the two responses confirms the selectivity of the edge state response.}
\end{figure}

The amplitude of the velocity spectra of two local areas (area 1: 3-row away from the DW; area 2: right on the DW) are plotted in Fig. \ref{fig:spectrum}. The large difference visible between the two responses, at approximately 17.8 kHz, indicates the edge state response.

\begin{figure*}[ht]
	\includegraphics[scale=1]{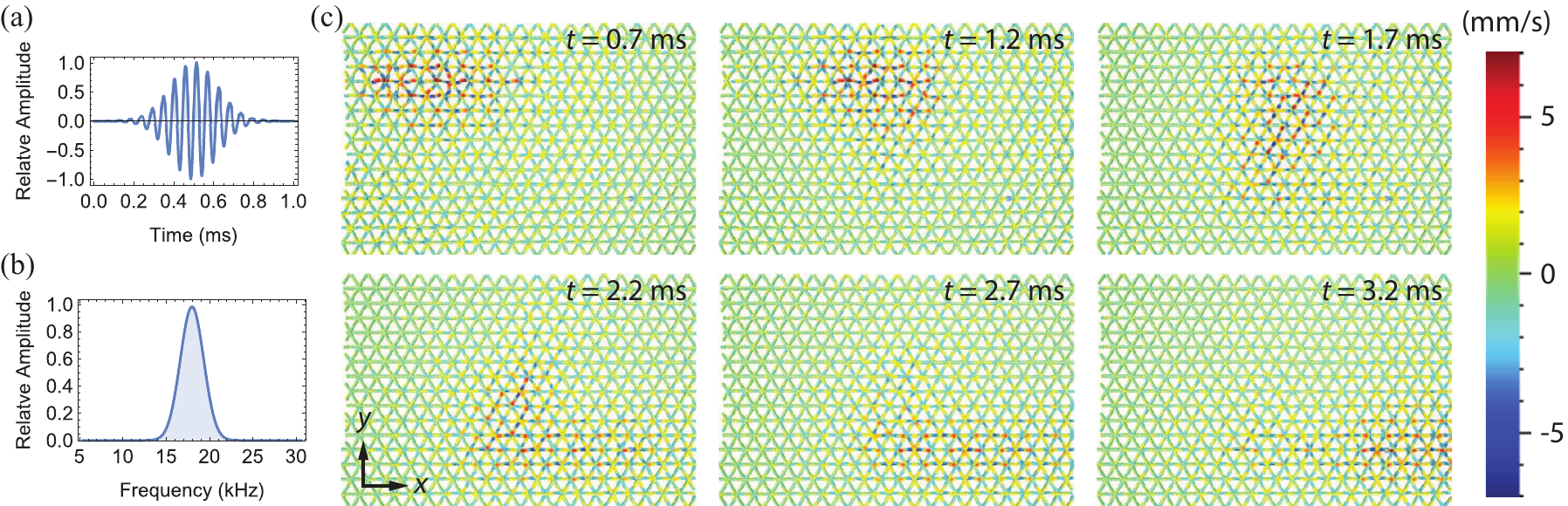}
	\caption{\label{fig:time} Measured transient response of a Gaussian sine pulse input. (a) The Gaussian sine pulse as the input signal. (b) The spectrum of the input signal. (c) Selected time instants are provided to show the propagation of the pulse along the DW. Despite the presence of a bulk mode (triggered during the initial transient), no reflections of the edge state are visible as the wave travels through the two corners of the Z-shaped DW.}
\end{figure*}

In order to further investigate the robustness of the edge states (i.e. the topological protection against back-reflections), we performed a time transient measurement. A Gaussian sine pulse (depicted in Fig. \ref{fig:time} (a)) with -3dB bandwidth in the range 16.4 to 19.6 kHz (Fig. \ref{fig:time} (b)) was used as actuation signal to the MFC actuator. Selected time instants from the measured response are shown in Fig.~\ref{fig:time}. Note that, in order to shorten the physical length of the pulse so to be entirely contained in each segment of the DW, the resulting frequency band was larger than the topological band gap.
This condition resulted in the excitation of bulk waves, that are visible particularly in the first time instant. Nevertheless, given the different wave speeds and propagation directions the different modes separate quickly therefore making the edge state very visible.
The time sequence clearly shows that the edge state does not reflect off the corners while traveling along the Z-shaped DW. A more intuitive interpretation of these results can be obtained based on the supplementary material \citep{supp}.

\begin{figure*}[ht]
	\includegraphics[scale=1]{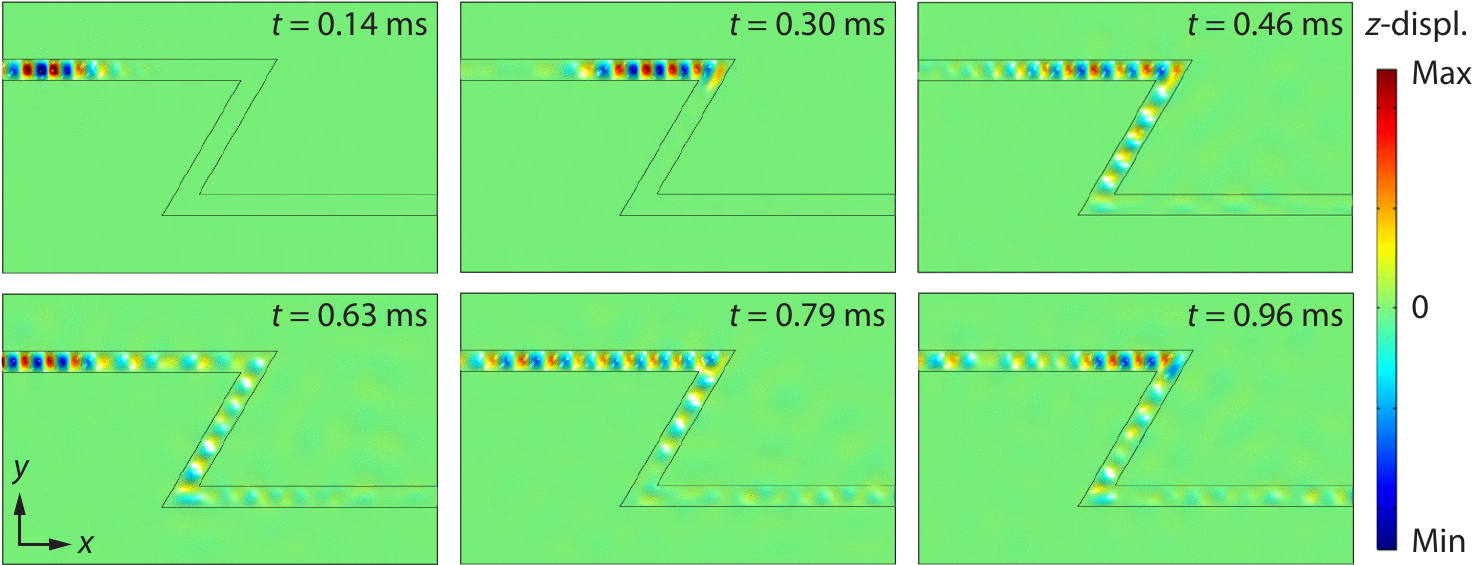}
	\caption{\label{fig:reflectionZ.pdf} Sequence extracted from the transient response of an ordinary (i.e. non-topological) Z-shaped waveguide. The strong backscattering from the 60 degree corner is evident and in net contrast with the topological response.}
\end{figure*}

In order to provide further evidence of the largely suppressed back reflection, we conducted a time-domain numerical analysis of a Z-shaped ordinary waveguide made of aluminum embedded in a steel background. From a geometric perspective, this Z-shaped waveguide was equivalent to the topological waveguide. The transient analysis shows unequivocally a strong backscattering from the 60 degree corner, hence drawing a net contrast with the topological waveguide.
Fig. \ref{fig:reflectionZ} shows snapshots taken at different time instants illustrating the reflection. The entire video of the transient response was uploaded as electronic supplemental material \cite{supp}.

\subsection{Selective valley injection}
\begin{figure}[h]
	\includegraphics[scale=2]{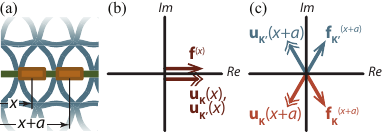}
	\caption{\label{fig:vinj} 
Illustration of the valley injection procedure. (a) Two identical actuators are installed on the DW at two locations separated by a distance $a$. The phase between the actuator input and the displacement at the two points can be schematically visualized in a complex plane. (b) Choosing the location $x$ as the reference, we can assume that the force $\mathbf{f}^{(x)}$ (the single arrow) and the corresponding particle displacements $\mathbf{u_K}(x)$ and $\mathbf{u_{K'}}(x)$ (the double arrow) are initially in phase. (c) The phase of the \textbf{K}- and $\mathbf{K}'$-polarized modes (the red and blue double arrows, respectively) and that of the harmonic forces at $x+a$ necessary to create pure \textbf{K}- and $\mathbf{K}'$-polarized modes (the red and blue single arrows, respectively). The forces produce negative work on the modes with different subscripts, therefore canceling the work done by the force at $x$ and realizing single-mode excitation.}
\end{figure}

Finally we show that, despite TRS is intact, the two valley-dependent edge states can be selectively excited in order to achieve unidirectional excitation. Consider a \textbf{K}-polarized edge mode ($k_xa=-2\pi/3$, see Fig.~\ref{fig:DW} (b)), from Bloch's theorem we know that for any two points in physical space separated by $a$ (for example, $x$ and $x+a$, see Fig.~\ref{fig:vinj} (a)) the phase of the particle displacement at $x+a$ differs from that at $x$ by $-2\pi/3$. Equivalently, a phase difference of $+2\pi/3$ is obtained if considering a $\mathbf{K}'$-polarized edge mode. Fig.~\ref{fig:vinj} (b) and (c) show a schematics of the displacement vectors (double arrows) at position $x$ and $x+a$ in a complex plane.

Based on the above considerations, we could use two harmonic point excitations located at $x$ and $x+a$ and actuated according to a prescribed phase difference related to the specific valley in which the injection is sought. For example, let the point excitation at $x+a$ have a phase $-\pi/3$ difference from that at $x$ (as the single red arrow $\mathbf{f_K}$ shown in Fig.~\ref{fig:vinj} (c)). The two-point source excites only a \textbf{K}-polarized edge mode, because it injects zero net energy to the $\mathbf{K}'$-polarized edge mode (equal amount but opposite signs, at points $x$ and $x+a$, respectively). Similarly a two-point source having phase difference $+\pi/3$ excites the $\mathbf{K}'$-polarized edge mode only (see the blue single arrow in Fig.~\ref{fig:vinj} (c)). An extensive description of the procedure can be found in \cite{myqvhe}.

\begin{figure}[h]
	\includegraphics[scale=0.84]{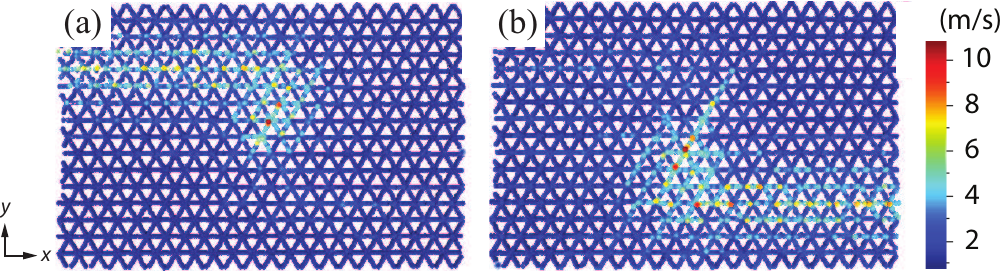}
	\caption{\label{fig:oneway}
	Experimental results showing the performance of the valley injection procedure. Targeted excitation of the uni-directional edge states is achieved by two harmonic sources with phase difference (a) $-\pi /3$ and (b) $+\pi /3$.}
\end{figure}

In the experiment, two identical MFC actuators separated by a distance $a$ were glued on the DW (see label Src.~set 2 in Fig.\ref{fig:setup} (a),(c)).
Fig~\ref{fig:oneway} (a) and (b) show the velocity amplitude response, resulting from an input phase difference of $-\pi/3$ and $+\pi/3$, respectively. The input frequency was fixed at 17.8 kHz.

\begin{figure}[h]
	\includegraphics[scale=0.45]{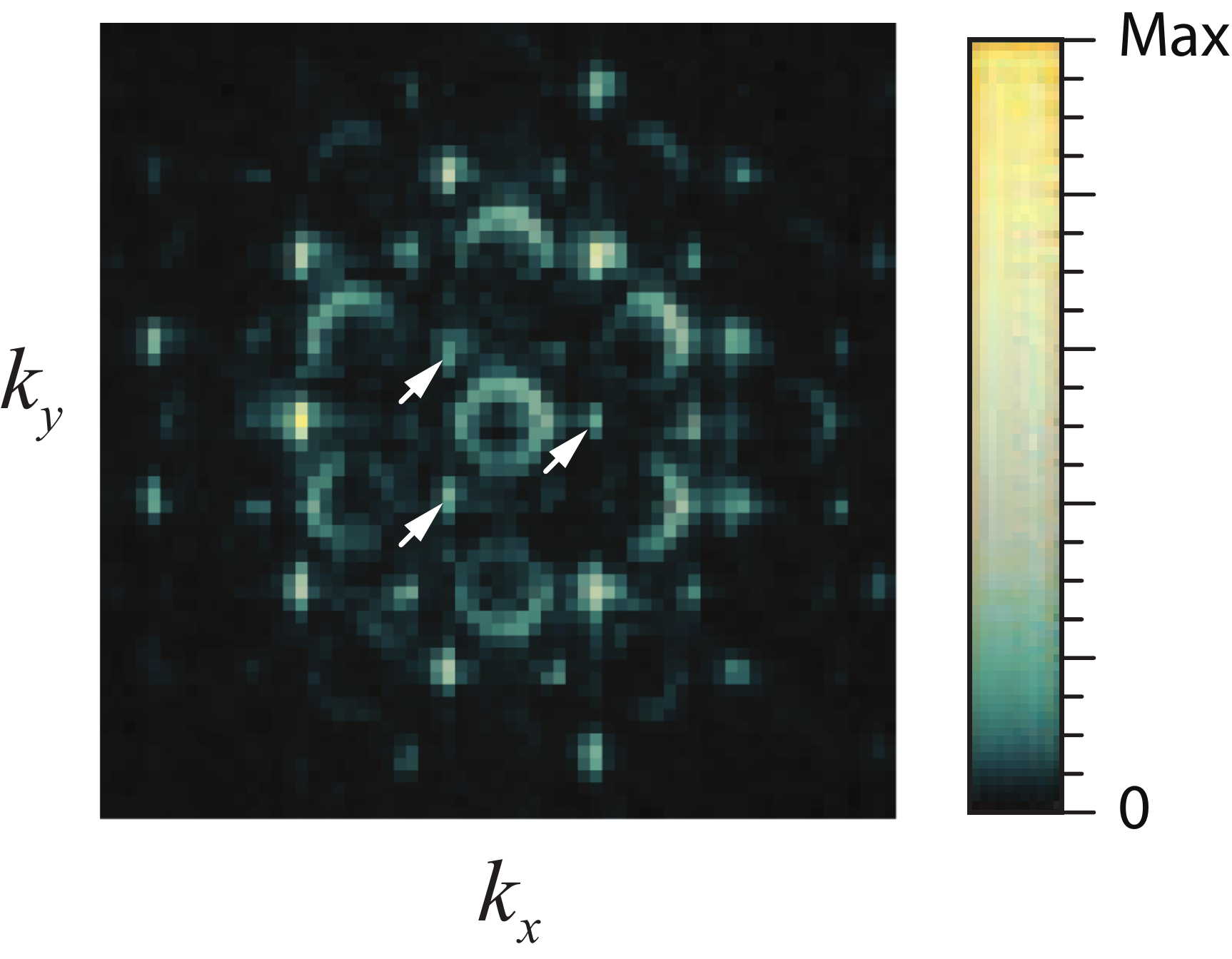}
	\caption{\label{fig:FFT}	
The Fourier spectrum of the velocity response corresponding to the \textbf{K} valley injection (Fig.~\ref{fig:oneway} (a)). Only the points corresponding to the \textbf{K} valleys and their reciprocal lattice duplication show high amplitude response. This result confirms accurate valley injection on the DW. The low-amplitude circles are due to the excitation of the bulk mode.}
\end{figure}

The experimental results show clear evidence of unidirectional excitation and further confirm the robustness of the edge states across discontinuities, since no appreciable response is visible in the remaining half of the DW.
The successful valley injection can also be confirmed by taking
the Fourier spectrum of the velocity response presented in Fig.~\ref{fig:oneway} (a). By visually inspecting the spectrum (Fig.~\ref{fig:FFT} (a)) the points corresponding to the valleys (and its reciprocal lattice duplication) exhibit have the highest velocity amplitude. The white arrows indicate the three $\mathbf{K}$ points on the boundary of the first BZ. The low-amplitude circular profiles are due to the presence of the bulk mode. A similar pattern would be obtained for data in Fig.~\ref{fig:oneway} (b). 
This excitation scheme can have remarkable engineering significance, not only because it is simple to implement but also because it can be applied universally to all AVHE systems.

Note that, in our experiment, we used MFC$^{\textregistered}$ thin film actuators. MFCs are 300$\mu$m thin interdigitated piezo fiber transducers that are very flexible in bending (similar to traditional strain gauges) and have low in-plane stiffness (E = 30GPa) compared with the supporting aluminum plate. These transducers were selected to minimize the mechanical impedance mismatch at the actuation point.

\section{Chiral vortices of energy flux}

\begin{figure*}[ht]
	\includegraphics[scale=1]{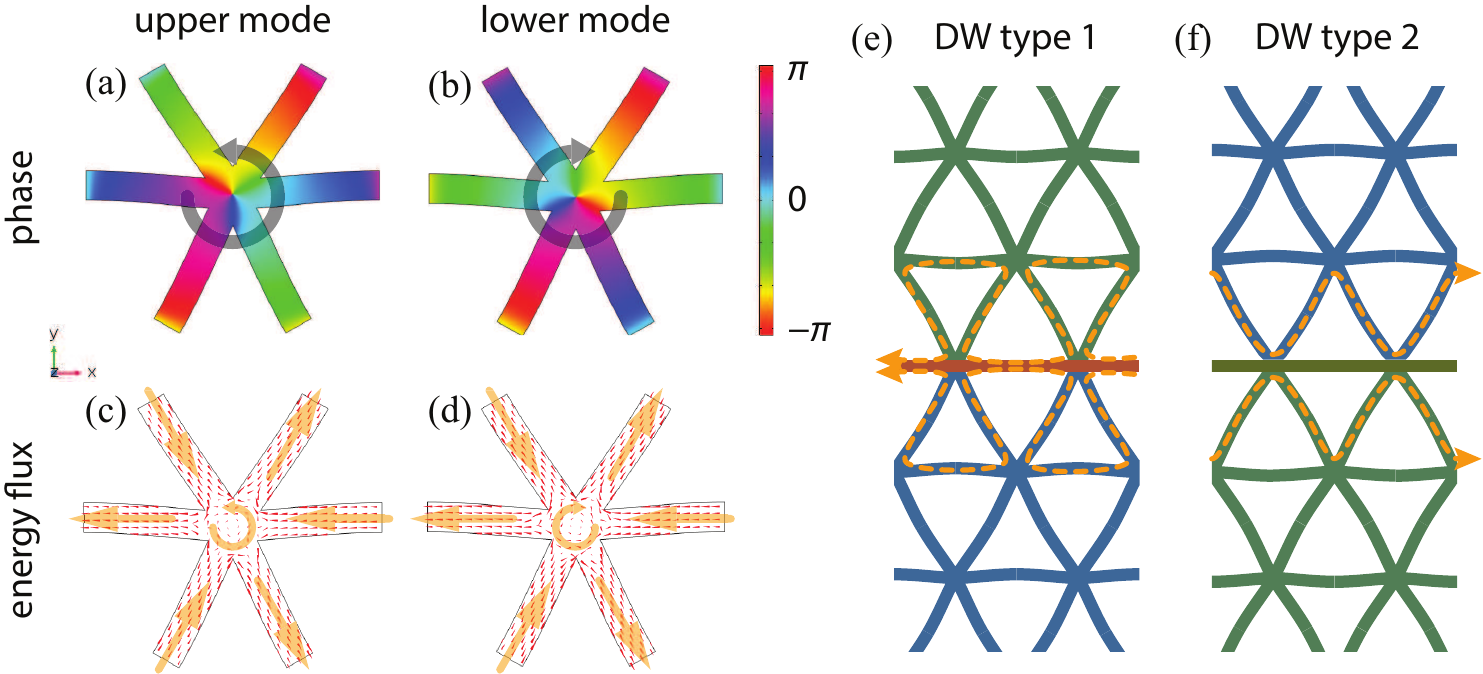}
	\caption{\label{fig:spin}
Numerical results showing the chiral vortex of the energy flux of the two modes at the \textbf{K} point of lattice B. (a, b) The phase of the $z$-component of the displacement field, and (c, d) the energy flux, for the upper mode (a, c) and the lower mode (b, d), respectively.}
\end{figure*}

\begin{figure}[h]
	\includegraphics[scale=0.75]{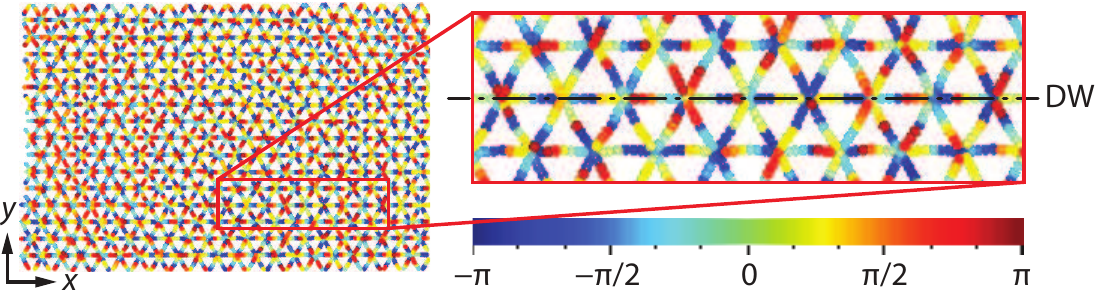}
	\caption{\label{fig:exp_phase}	
The phase distribution extracted from the experimental measurements corresponding to Fig. \ref{fig:oneway} (b). The close-up view near the DW shows that the phase distribution is symmetric with respect to the DW, therefore leading to counter-circulating flux within the lattices separated by the DW.}
\end{figure}

In this section, we perform an in depth analysis of the eigenmodes in order to provide additional physical insight into the robustness of the edge states.
Although the local topological nature of the material is already indicated by the valley Chern numbers, a further indication of the topologically non-trivial nature of the medium is confirmed by the existence of chiral edge states supported in the topological bandgap at the valleys. Taking lattice B as the example, Fig.~\ref{fig:spin} shows the phase of the $z$-component (the most significant component associated with the flexural mode) of the displacement field (a, b), and the corresponding energy flux (c, d) associated with the two nondegenerate modes at the \textbf{K} point (namely, the upper mode (a, c) and the lower mode (b, d)). 

Observing the phase distribution in Fig.~\ref{fig:spin}(a,b), we note the presence of a branch point at the center of the unit cell (i.e. the intersection of the trusses) either for the upper or lower modes. Each branch point is characterized by opposite phase winding numbers that take the value $+1$ or $-1$.
This parameter represents the number of complete turns (module $2\pi$) of the phase change accumulated along a small closed loop enclosing the singular point, for a fixed time instant.

In addition, similar energy flux patterns are observed for both the upper and lower \textbf{K}-polarized modes (see Fig.~\ref{fig:spin} (c) and (d)). Clockwise energy flux circulates in each upright triangle (another possible choice for the primitive unit cell). 

Concerning the other two modes at the $\mathbf{K}'$ point, they possess opposite chirality for both the phase and the energy flux distributions. This is a direct consequence of TRS being intact.
Based on the above observations, it is suggested that the sign of the winding number around the inversion center in physical space is connected to the sign of the valley Chern number. Also, given a topological lattice and a selected valley point, the eigenmodes associated to the edge states exhibit valley-dependent chiral vortices of the energy flux.
Modes with opposite chirality supported by different valleys do not couple unless in presence of defects that strongly perturbs the chirality of the energy flux. It is merely suggested here that the valley index ($\mathbf{K/K}'$) in our lattice are reminiscent, from a purely qualitative point of view, of the spin index in the QSHE topological insulators.

Obviously, the valley and spin indices are two independent quantities connected to separate mechanisms. This difference is also reflected in their mathematical modeling when comparing the terms introduced into the corresponding Dirac Hamiltonian \cite{ReviewKaneTI}. However, we observe that the direction of the energy flux vortices in our system is connected to the valley index. It follows that back-scattered modes are less likely to happen (assuming inter-valley hopping is prohibited) for a valley-polarized mode unless the back-scattering process can produce a change in the energy chirality. This mechanism can be further clarified by the fact that the a back-scattered edge mode is the time-reversed counter part of the incident (forward propagating) mode, therefore the reflection process would require also a sign inversion of the valley index (hence of the energy vortex chirality). This idea recalls, from a very qualitative perspective, the condition occurring in QSHE systems in which the reflection of edge states requires inversion of the spin index (which, being prohibited, leads to robust edge states).

In addition, consider the energy flux associated with the edge states. Figures~\ref{fig:spin} (e) and (f) provide a schematic view of the energy flux for the \textbf{K}-polarized edge state ($k_xa=-2\pi/3$) in the lattice cells neighboring to the interface (orange dashed lines). Although the energy flux direction in each truss is similar to those of the \textbf{K}-polarized bulk states (shown in Fig.~\ref{fig:spin} (c, d)), in the edge states the energy flux does not give rise to complete circulations, but instead travels along meandering paths along the DW. This mechanism is reminiscent of the semi-classical interpretation of the integer QHE where the magnetic field induces a cyclotronic motion of the electrons in the bulk (the insulating phase), and a typical \textit{skipping orbit} motion of the electrons along the edges (which gives rise to the uni-directional edge states). Similarly in our elastic AVHE, the bulk is characterized by a locally circulatory energy flux which result in an \textit{acoustically insulating} material, while the edge states are formed and can propagate uni-directionally due to the interruption of the energy flux circulation at the DW. Note that, unlike QHE materials in which the existence of gapless edge state is allowed in a single nontrivial insulator, the AVHE requires assembling two materials with opposite valley Chern number (or, equivalently, single material with selected boundary conditions) \cite{myqvhe}.

Fig. \ref{fig:exp_phase} shows the phase response extracted from the same measurement of Fig. \ref{fig:oneway} (b). The close-up view near the DW shows that the phase distribution is symmetric with respect to the DW, therefore leads to counter-circulating directions in the two lattices separated by the DW.

\section{conclusions}

This experimental study presented evidence of the existence and robustness of topological edge states in non-resonant phononic elastic waveguides, hence extending the applicability of the AVHE concept beyond the conventional locally-resonant phononic crystals.
The test sample consisted in a reticular aluminum thin plate designed according to the acoustic analogue of the quantum valley Hall effect. Edge states were obtained by contrasting two slabs of the lattice having broken and inverted SIS.
The experimental data provided strong evidence of the existence of the edge states and, more interestingly, of a highly suppressed inter-valley mixing. This latter characteristic is of extreme importance to achieve unidirectional waveguides. In fact, even if TRS is intact and reciprocal modes are simultaneously supported by the domain wall, the weak coupling between edge modes strongly suppresses the scattering of the edge states when interacting with discontinuities. Such behavior was also further supported by the numerical and experimental observation of a chiral flux of energy whose direction is determined by the valley polarization. The two lattices forming the DW show opposite directions of chirality (from an energy flux perspective) which is one of the effects contributing to the suppressed back-scattering. These chiral vortices of the energy flux also share interesting qualitative similarities with the inter-valley hopping and the skipping orbit motion characteristic of the quantum Hall effect.
Finally, this study also provided experimental evidence in support of a selective-valley injection method which enables accurate excitation of unidirectional edge states.

\begin{acknowledgments}
The authors gratefully acknowledge the financial support of the Air Force Office of Scientific Research under Grant No. YIP FA9550-15-1-0133. The authors also thank Mr. Janav Udani and Dr. Andres Arrieta for help with the electronic equipment.

\end{acknowledgments}

\bibliography{ref2}

\end{document}